# Polar jets of swimming bacteria condensed by a patterned liquid crystal

Taras Turiv[1], Runa Koizumi[1], Kristian Thijssen[2], Mikhail M. Genkin[3], Hao Yu[1], Chenhui Peng[1], Qi-Huo Wei[1,4], Julia M. Yeomans[2], Igor S. Aranson[5], Amin Doostmohammadi[2], Oleg D. Lavrentovich[1,4]*

[1]Advanced Materials Liquid Crystal Institute, Chemical Physics Interdisciplinary Program, Kent State University, Kent, OH 44242, USA

[2]The Rudolf Peierls Centre for Theoretical Physics, University of Oxford, Clarendon Laboratory, Parks Rd., Oxford, OX1 3PU, UK

[3]Cold Spring Harbor Laboratory, 1 Bungtown Rd, Cold Spring Harbor, NY 11724, USA

[4]Department of Physics, Kent State University, Kent, OH 44242, USA

[5]Department of Biomedical Engineering, Pennsylvania State University, University Park, Pennsylvania 16802, USA

*Author for correspondence: e-mail: olavrent@kent.edu, tel.: +1-330-672-4844

## Abstract

Active matter exhibits remarkable collective behavior in which flows, continuously generated by active particles, are intertwined with the orientational order of these particles. The relationship remains poorly understood as the activity and order are difficult to control independently. Here we demonstrate important facets of this interplay by exploring dynamics of swimming bacteria in a liquid crystalline environment with pre-designed periodic splay and bend in molecular orientation. The bacteria are expelled from the bend regions and condense into polar jets that propagate and transport cargo unidirectionally along the splay regions. The bacterial jets remain stable even when the local concentration exceeds the threshold of bending instability in a non-patterned system. Collective polar propulsion and different role of bend and splay are explained by an advection-diffusion model and by numerical simulations that treat the system as a two-phase active nematic. The ability of prepatterned liquid crystalline medium to streamline the chaotic movements of swimming bacteria into polar jets that can carry cargo along a predesigned trajectory opens the door for potential applications in cell sorting, microscale delivery and soft microrobotics.

## Introduction

Active matter, an out-of-equilibrium assembly of moving and interacting objects that locally convert energy into directed motion, shows an intriguing interplay of dynamics and spatial order (*1*). This order is orientational, naturally associated with the vector character of velocity and with the elongated shape of moving units such as motile bacteria. A conventional (passive) nematic liquid crystal (LC) exhibits a long-range orientational order along an axis of anisotropy called the director $\hat{\mathbf{n}} \equiv -\hat{\mathbf{n}}$. A "living nematic" can be formed by adding swimming bacteria (*2*) (Fig. 1). The director, either uniform (*2-6*), or spatially distorted (*2, 6-9*), serves as an easy swimming pathway for bacteria. A living nematic allows one to control independently the activity, through concentration and speed of bacteria (*2, 10*) and orientational order, through predesigned director patterns (*6*)·(*7*).

The interconnection of activity and director is clearly expressed by the active force, introduced by Simha and Ramaswamy (*11*), $\mathbf{f}_a = \alpha \left( \hat{\mathbf{n}} \nabla \cdot \hat{\mathbf{n}} - \hat{\mathbf{n}} \times \nabla \times \hat{\mathbf{n}} \right)$, where $\alpha$ is the activity coefficient, and the director gradients represent splay $\hat{\mathbf{n}} \nabla \cdot \hat{\mathbf{n}}$ and bend $\hat{\mathbf{n}} \times \nabla \times \hat{\mathbf{n}}$ (*1, 7, 11, 12*) (Fig. 1). The force $\mathbf{f}_a$ causes instabilities, such as bending of flow trajectories and nucleation of topological defects (*2, 11, 13, 14*). However, when the living nematic (or other active matter, such as a dispersion of active microtubules (*15-17*)) is in a controlled confinement, $\mathbf{f}_a$ can be pre-imposed through surface alignment of $\hat{\mathbf{n}}$, to play a constructive role in commanding the dynamics (*6, 7, 12*). For example, a hybrid alignment (tangential at one plate and perpendicular at the other plate) of a flat cell with a living nematic rectifies the bacterial motion (*6*), while a splay-bend vortex causes a polar circulation of bacteria (*7*). These effects manifest a general prediction by Green et al. that a threshold-less macroscopic flows arise whenever $\nabla \times \mathbf{f}_a \neq 0$ (*12*). Yet the relationship between the director gradients and the activity is far from being well understood. One intriguing issue is that the concentration of active elements is not necessarily uniform in space. In particular, in the presence of topological defects, the bacteria concentrate near defects of a positive topological charge and avoid those with negative charges (*7, 8*). Furthermore, the form of $\mathbf{f}_a$ suggests some parity of the splay and bend, but it is not clear whether this parity is preserved when the concentration $c$ of bacteria (and thus $\alpha$, which is proportional to $c$) varies in space; most theories deal with incompressible active matter, $\alpha = \text{const}$, $c = \text{const}$. Maitra et al. (*18*) suggested recently that thin cells of living nematics might impart a different "weight" to the splay and bend contributions to $\mathbf{f}_a$.

In this work, in order to understand the interplay of activity, director gradients, and spatially varying concentration of active units, we explore a living nematic that is predesigned into a periodic pattern of alternating splay and bend. The director pattern forces a strongly non-uniform distribution of swimming bacteria, condensing them into polar "jets" that move unidirectionally along the maximum splay. The bend regions are mostly deprived of the swimmers and serve (i) to reverse the polarity of bacterial trajectories, (ii) to expel the bacteria towards splay regions and (iii) to stabilize the shape of the condensed jets. The jets retain a stable straight shape even when the concentration of bacteria in them exceeds (by an order of magnitude) the threshold of bend instability observed in a uniformly aligned cell. The concentration gradients and jets are described by an analytical advection-diffusion model that operates with two subsets of bacteria swimming in opposite directions (*8*). We further complement the experiments with numerical simulations of a two-phase model of active nematics to show that the difference between the bacterial response to splay and bend can be attributed to a generic mechanism of active concentration exchange between the two regions and to demonstrate the enhanced stability of the condensed jets against bending instabilities.

**Results and discussion**

The living nematic is represented by a water-based chromonic nematic disodium cromoglycate (DSCG) with dispersed bacteria *Bacillus subtilis*. We explore diluted, (spatially averaged concentration $\langle c \rangle \approx 0.3 \times 10^{14} \ \mathrm{m^{-3}}$) and concentrated ($\langle c \rangle \approx 1.5 \times 10^{14} \ \mathrm{m^{-3}}$) dispersions. The living nematic is confined by two glass plates separated by a gap $d = 20 \ \mu\mathrm{m}$ (Fig. 1a). The surfaces of the two plates are photoaligned to produce identical director patterns, as described previously (*7, 19*). To facilitate the presentation, we consider the director of the passive LC as a vector $\hat{\mathbf{n}}$, which is permissible in the absence of disclinations (*20*). The photopatterned vector field written in Cartesian coordinates as

$$\hat{\mathbf{n}}(x, y, z) = \big[\cos(\pi y / L), -\sin(\pi y / L), 0\big], \tag{1}$$

represents a one-dimensional modulated splay and bend in the $xy$-plane of the cell (Fig. 1b,c); $L = 160 \ \mu\mathrm{m}$. The pattern resembles rows of letters "c". Pure splay deformations are located at $y = 0, \pm L, \pm 2L, \ldots$, while bend at $y = \pm L / 2, \pm 3L / 2, \ldots$. The corresponding active force

$$\mathbf{f}_a = \alpha \pi / L \left[ -\cos\left(2\pi y / L\right), \sin\left(2\pi y / L\right), 0 \right], \qquad (2)$$

assuming $\alpha = \text{const}$ and $\alpha < 0$ (swimming bacteria are pushers), is plotted in Fig. 1d; it is expected to trigger flows towards the positive ends of the $x$-axis in the splay regions and in the opposite direction in the bend regions. However, as detailed below, the assumption $\alpha = \text{const}$ and the parity of splay and bend are at odds with the experiment, since the bacteria show a much higher affinity to splay than to bend.

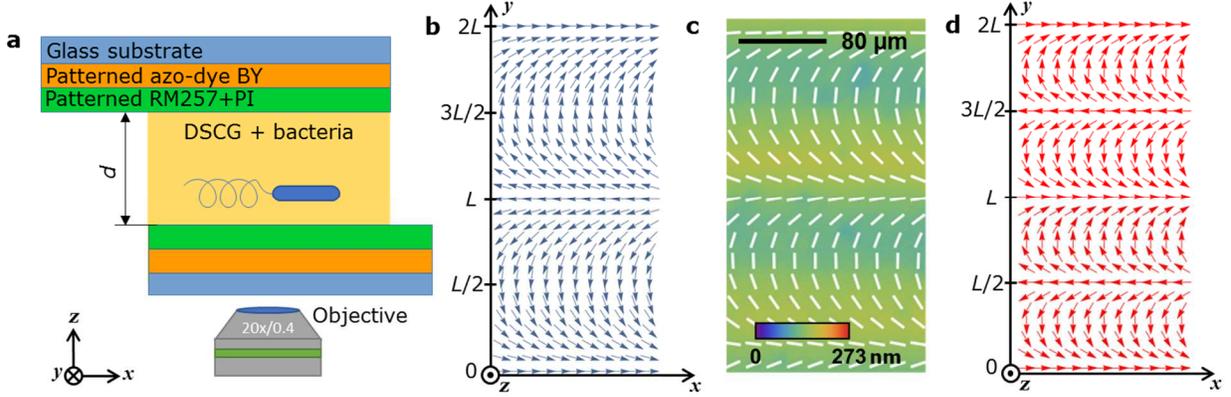

**Fig. 1**. Experimental scheme of the cell with bacteria swimming in a patterned nematic DSCG. **a**, Schematics of the experimental setup and LC cell. **b**, Predesigned director field. **c**, Optical phase retardation (background colors) and director field of the photopatterned nematic phase at $24°C$ mapped by the PolScope. **d**, Spatial variation of the active force $\mathbf{f}_a$ for extensile swimmers, arising from splay and bend of the director in equation (2), assuming $\alpha = \text{const}$.

_**Weakly concentrated dispersion.**_ Once the dispersion is injected into the patterned cell, the swimming bacteria start to leave the bend regions and accumulate in the splay regions, moving predominantly along the positive direction of the $x$-axis (Fig. 2a,b). The experiment demonstrates a dramatic difference between the splay and bend regions, as there are no net flows towards the negative end of the $x$-axis in the bend regions. The bacterial concentration $c$, calculated by averaging over the $x$-axis, is strongly modulated along the $y$-axis; time evolution of $c(y)$ is shown in Fig. 2c. The distribution $\langle c(y) \rangle_t$, averaged over 10 min, is of the same period $L$ as the splay-bend pattern; in the splay regions, $c(y)$ reaches $\approx 4.4 \times 10^{14}$ m$^{-3}$ (Fig. 2d), 15 times higher than the initial concentration $\langle c \rangle \approx 0.3 \times 10^{14}$ m$^{-3}$.

The velocity field of bacteria is biased towards the converging side of splay (Fig. 2e,f and Supplementary Video 1). If an individual bacterium happens to swim in the opposite direction, the diverging director turns it away from the splay and forces it to make a U-turn by swimming along the director in the bend region and to join a bacterial jet moving along the $x$-axis in the neighboring splay region. The velocity components $v_x(y)$ and $v_y(y)$ are strongly modulated along the $y$-axis (Fig. 2g); $v_x(y)$ is predominantly positive, reflecting polar transport along the $x$-axis, while $v_y(y)$ fluctuates near zero, demonstrating absence of net flows along the $y$-axis.

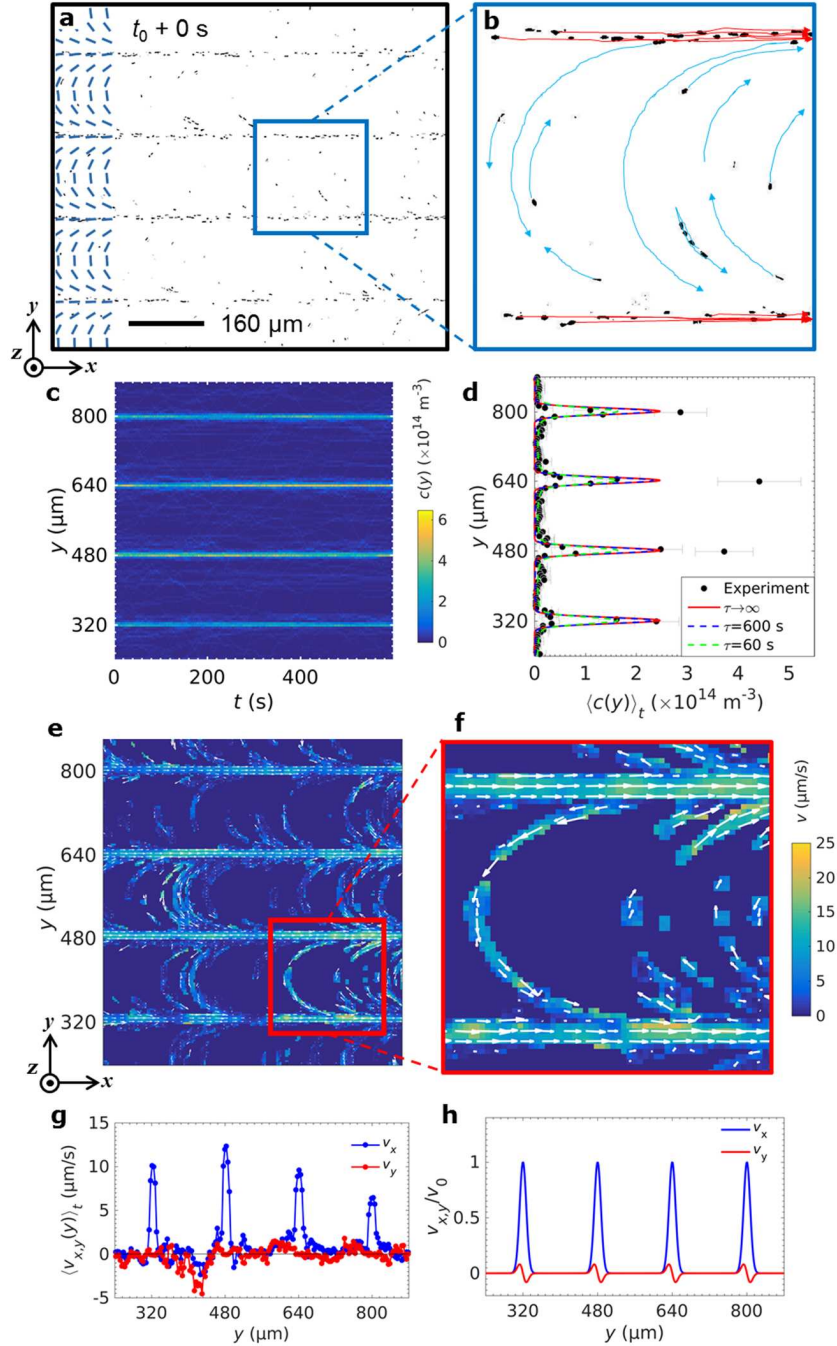

**Fig. 2.** Trajectories, distribution and velocities of the bacteria in a weakly concentrated dispersion, $\langle c \rangle \approx 0.3 \times 10^{14}$ m$^{-3}$, controlled by a periodic splay-bend director pattern of DSCG. **a,b,** Enhanced (by subtracting the average image over all frames of the video) microscopic images of bacteria

condensed in the splay regions; part **b** shows trajectories of bacteria moving in splay (red curves) and crossing the bend (blue curves with the arrows indicate the direction of motion). **d,** Kymograph of the concentration $c(y)$ with the time step 0.5 s. **e,** Time average of bacterial concentration $\langle c(y) \rangle_t$ (errors are standard deviations of concentration), and theoretical curves $c(y)$ calculated from equation (6) with $v_0 = 10\,\mu\text{m/s}$, $D_c = 10\,\mu\text{m}^2/\text{s}$, $L = 160\,\mu\text{m}$, and $C_0 = 1.9 \times 10^{-8}\,\text{m}^{-3}$ and for an infinite $\tau$; also shown are the results of advection-diffusion simulations for finite reversal times, $\tau = 600\,\text{s}$ and $\tau = 60\,\text{s}$. **e,f,** Experimental velocity map; the length of arrows and the pseudocolors scale linearly with the speed of bacteria. **g,** Experimental velocity components $v_x(y)$ and $v_y(y)$. **h,** Velocity components, weighted by the concentrations of bacteria swimming parallel $c^+$ and antiparallel $c^-$ to vector $\hat{\mathbf{n}}$, $v_{x,y} = v_{x,y}^+ c^+ + v_{x,y}^- c^-$, obtained from the advection-diffusion simulation, normalized by the maximum velocity $v_0$. All data are averaged over 600 s. See Supplementary Video 1.

Condensation of bacteria in the splay regions and formation of polar jets follows from the strong preference of bacteria to swim along the director. Swimming along any other direction involves penalties associated with increased viscous resistance, bulk elastic and surface anchoring energy cost, as discussed in Ref. (*21*) and in the Supplementary Information (SI). To describe the effect, we employ the model proposed by Genkin et al (*8*), in which transport of bacteria in a LC is governed by two coupled advection-diffusion equations for the concentrations $c^\pm$ of bacteria swimming parallel ($c^+$) or anti-parallel ($c^-$) to the vector $\hat{\mathbf{n}}$,

$$\partial_t c^\pm + \nabla \cdot \left( \pm v_0 \hat{\mathbf{n}} c^\pm + \mathbf{v}_f c^\pm \right) = \mp \frac{c^+ - c^-}{\tau} + D_c \nabla^2 c^\pm; \tag{3}$$

here $v_0 \sim 10\,\mu\text{m/s}$ is the bacterium swimming speed, $D_c$ is the concentration diffusion coefficient, $\mathbf{v}_f$ is a flow velocity, and $\tau$ is the direction reversal time. A direction reverse means that the bacterium leaves the population $c^+$ and joins $c^-$, or vice versa. Our experiments show that $\tau \sim 10^4$ s (SI and Supplementary Video 1) is much larger than the typical duration of the experiment. In the tumbling-free $\tau \to \infty$ limit and for $\mathbf{v}_f = 0$ (fluid velocity is small compared to the propulsion velocity), the equations for $c^\pm$ decouple; integrating these stationary equations (3), one obtains

$$v_0 n_y c^\pm = D_c \partial_y c^\pm + C_{1,2}. \tag{4}$$

We set constants $C_{1,2}$ to zero in order to obey the periodicity condition. Then the solutions are

$$c^{\pm} = C^{\pm} \exp\left(\mp \frac{v_0 L}{\pi D_{\mathrm{c}}} \cos \frac{\pi y}{L}\right), \qquad (5)$$

here $C^{+} = C^{-} = C_0 / 2$ (see SI), and $C_0$ is determined from the condition $\frac{1}{H}\int c(y)\,dy = \langle c\rangle$, where $\langle c\rangle \approx 0.3 \times 10^{14}\ \mathrm{m}^{-3}$; $H \approx 10^3\ \mu\mathrm{m}$ is the length of patterned region along the $y$-axis. The resulting $y$-dependence of the concentration $c = c^{+} + c^{-}$,

$$c(y) = C_0 \cosh\left(\frac{v_0 L}{\pi D_{\mathrm{c}}} \cos \frac{\pi y}{L}\right), \qquad (6)$$

is plotted in Fig. 2d, together with the dependencies $c(y)$ for two finite values of $\tau$, 60 s and 600 s, obtained by numerical integration of equation (3) following Ref. (8), see SI for details. Finite $\tau$ does not alter the qualitative behavior predicted by equation (6).

The analytical solution (6) and simulations thus explain the most salient feature of the experiment, namely, polar unidirectional flows in the splay regions and absence of a directed net flow in the bend regions (Fig. 2h). There are some quantitative discrepancies: the experimental spikes of concentration in Fig. 2d are 30%-40% higher and narrower than the peaks predicted by equation (6). There are two plausible reasons. First, bacterial diffusion perpendicular to $\hat{\mathbf{n}}$ is more difficult than along $\hat{\mathbf{n}}$. This diffusion anisotropy would make the theoretical peaks sharper, but it is not accounted for in the theory. Second, the bacteria could realign the patterned director closer to the $x$-axis in the vicinity of splay regions because of the director anchoring at their surface (21) which also would help to accumulate higher numbers of bacteria there. Note that in the computational model we operate with the continuous concentration fields, therefore the anchoring on bacterial surface cannot be accounted for.

***Undulation of bacteria jets for highly concentrated dispersion.*** An intrinsic feature of active matter is the emergence of bending and splay instabilities and nucleation of topological defects at elevated activities (2, 13, 22-24). Finding the means to suppress these instabilities and to convert activity into controllable steady flows might significantly expand the potential of active matter for applications. Below we demonstrate that the patterned splay-bend director stabilizes rectilinear bacterial flows against bending.

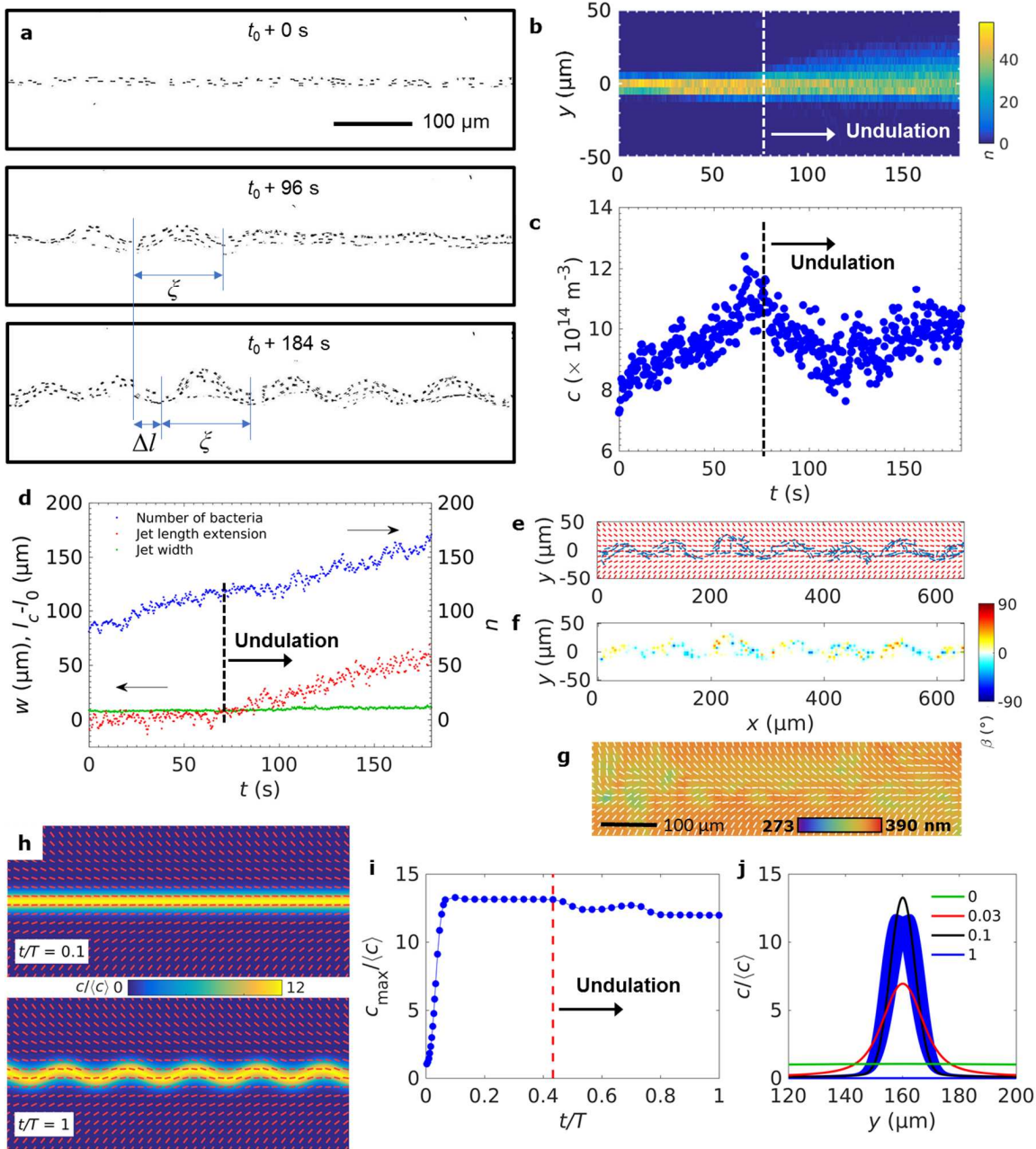

**Fig. 3.** Emergence of the bacteria jet undulations for highly concentrated dispersion, $\langle c \rangle \approx 1.5 \times 10^{14}$ m$^{-3}$. **a**, Enhanced microscopic images of bacteria at different stages of undulation development. The undulating wave translates from left to right with the speed $\approx 0.3$ μm/s. **b**, Kymograph of bacterial distribution along the $y$-axis with the time step 0.2 s; the color bar shows the number of bacteria per length step $\Delta y = 5$ μm. **c**, Time evolution of bacterial concentration in the developing jet. **d**, Total number of bacteria $n$, excess of the jet contour length $l_c - l_0$, where $l_0 = 650$ μm is the length of the rectilinear jet, and average width $w$ of the jet. **e**, Local director (red ticks) and bacteria body axes (blue arrows) for fully developed undulation at time $t_0 + 184$ s. **f**, Map of the angle $\beta$ between the surface-imposed director and bacteria body axes. **g**, Optical retardation and director map for an undulating instability imaged by PolScope. **h**, Advection-diffusion simulation of the director and normalized bacteria concentration for straight and undulating jets. **i**, Maximum bacteria concentration as a function of normalized computational time $t/T$, where $T$ is the maximum simulation time. **j**, The simulated distribution of bacterial concentration in the splay region at different computation times (shaded blue for undulated jet, since concentration is $x$-dependent, from $c_{max} = \max\left(c\left(x,y\right),x\right)$ to $c_{min} = \min\left(c\left(x,y\right),x\right)$) obtained from simulation following Ref. [8]. See SI for the details of the simulation parameters.

In uniform cells, increasing the bacterial concentration above $c_{u}^{uniform} \approx 0.9 \times 10^{14}$ m$^{-3}$ leads to a bending instability of their trajectories (Supplementary Fig. 1 and Supplementary Video 2). In patterned cells of the same thickness 20 μm, however, the bacterial jets start to undulate only at $c_{u}^{s\text{-}b} \approx 12 c_{u}^{uniform} \approx 11.8 \times 10^{14}$ m$^{-3}$; the speed of bacteria in the jets remains nearly constant at 10 μm/s below and above $c_{u}^{s\text{-}b}$ (Supplementary Fig. 2a). Interestingly, right after the undulations emerge, the contour length of the jet increases a bit faster than the number of bacteria in it, so that the concentration slightly decreases (Fig. 3c,d and Supplementary Fig. 2b). The stabilizing action of the splay-bend patterns is evident even at $c > c_{u}^{s\text{-}b}$ (Fig. 3a,b and Supplementary Video 3), since the wave of undulations preserves its wavelength, $\xi \sim 100$ μm, and amplitude (Fig. 3b). Unlike the bending instabilities in uniform cells (*2*), the undulations of jets do not grow to nucleate pairs of topological defects.

The stabilizing effect of the splay-bend pattern can be explained by the fact that an undulating jet tilts with respect to the surrounding director, causing an increase in the elastic and interfacial energy. To illustrate this, we calculate the angle $\beta$ between the axes of swimming bacteria and the local director imprinted at the substrates (Fig. 3e-g and Supplementary Fig. 2c-g). For the undulating jet, the standard deviation of $\beta$ from 0, characterizing perfect alignment along $\hat{\mathbf{n}}$, is $\Delta\beta = \sqrt{\sum_i \beta_i^2 / n} \approx 36°$, where $\beta_i$ is the angle measured for $n$ individual bacteria in each frame (Supplementary Fig. 2c). The large $\beta$ implies director twist along the $z$-axis, which is visualized by observations with PolScope (Fig. 3g and Supplementary Figs. 2,3), as regions with a diminished effective optical retardance, caused by the change of light polarization from linear to elliptical.

The numerical simulations based on the model by Genkin et al. (*8*), demonstrate that the undulating jets indeed distort the pre-imposed director (Fig. 3h and Supplementary Video 4). The concentration of bacteria inside the jet increases rapidly and remains constant until the undulating jet is formed (Fig. 3i) after which it slightly decreases, similarly to the experimental observations.

The distribution of bacteria concentration becomes wider after the establishment of the undulation (Fig. 3j). The undulating wave propagates in the same direction as the bacterial motion with a typical velocity of $0.5\,\mu m/s$, in agreement with the experimental observation (Supplementary Videos 3,4).

The dramatic difference in the concentration of bacteria in the splay and bend regions allows us to further extend the theoretical description by introducing a two-phase continuum model that captures the long-range hydrodynamic, collective flow interactions between the bacteria which result in the undulation. The two-phase model is better suited to describe strongly concentrated regimes as compared to the advection-diffusion model developed around a dilute limit. We define a phase-field order parameter $\phi$ such that $\phi = 1$ demarcates the active regions occupied by bacteria, where the nematic tensor $\mathbf{Q}$ has a finite value, while $\phi = 0$ corresponds to the isotropic passive regions void of any activity with $\mathbf{Q} = 0$. The dynamics of $\phi$ is governed by a Cahn-Hilliard equation (25), resulting in a diffuse interface between the active phase and the surrounding fluid (26) and it is coupled to the velocity field generated by active particles (see SI text).

An important contribution to the orientation dynamics is the alignment of bacteria along the underlying director pattern. To model this, we contain all bacteria-LC interactions in a Rapini-Papoular-like free energy term, which penalizes orientations of the active units that are not parallel to $\hat{\mathbf{n}}$. The preferred orientation of $\hat{\mathbf{n}}$ at the *B. subtilis* surface is tangential (21). Since the experiments are performed in a thin-film geometry, $d \ll L$, we use a quasi-2D approximation detailed in the SI. The analysis does not account for the twist deformations observed at concentrations well above $c_u^{s-b}$, since our prime interest here is in the behavior below and closely above the undulation threshold.

The simulation is initialized with a stripe of active extensile phase (pusher type bacteria) denoted by $\phi = 1$ placed along the $y$-direction in the splay-bend director pattern (Fig. 4a) and the rest of the system with $\phi = 0$. As the simulation evolves, the initially uniform active stripe focuses into the domain that migrates to the splay region, and this migration continues until the bend region is completely devoid of the active phase. Consistent with the experiment and the analytical advection-diffusion model, once the active phase is focused within the splay region, a unidirectional flow persists towards the converging side of splay causing strong modulation of the concentration (Fig. 4a,c and Supplementary Video 5).

In the simulation, the focusing is caused by the active forces converging towards splay. In the bend regions, the forces are diverging, helping to deplete these regions. Conversely, for a contractile active system, this synergy leads to depletion of splay regions and transport of active phase to the bend region (Supplementary Fig. 4 and Supplementary Video 6).

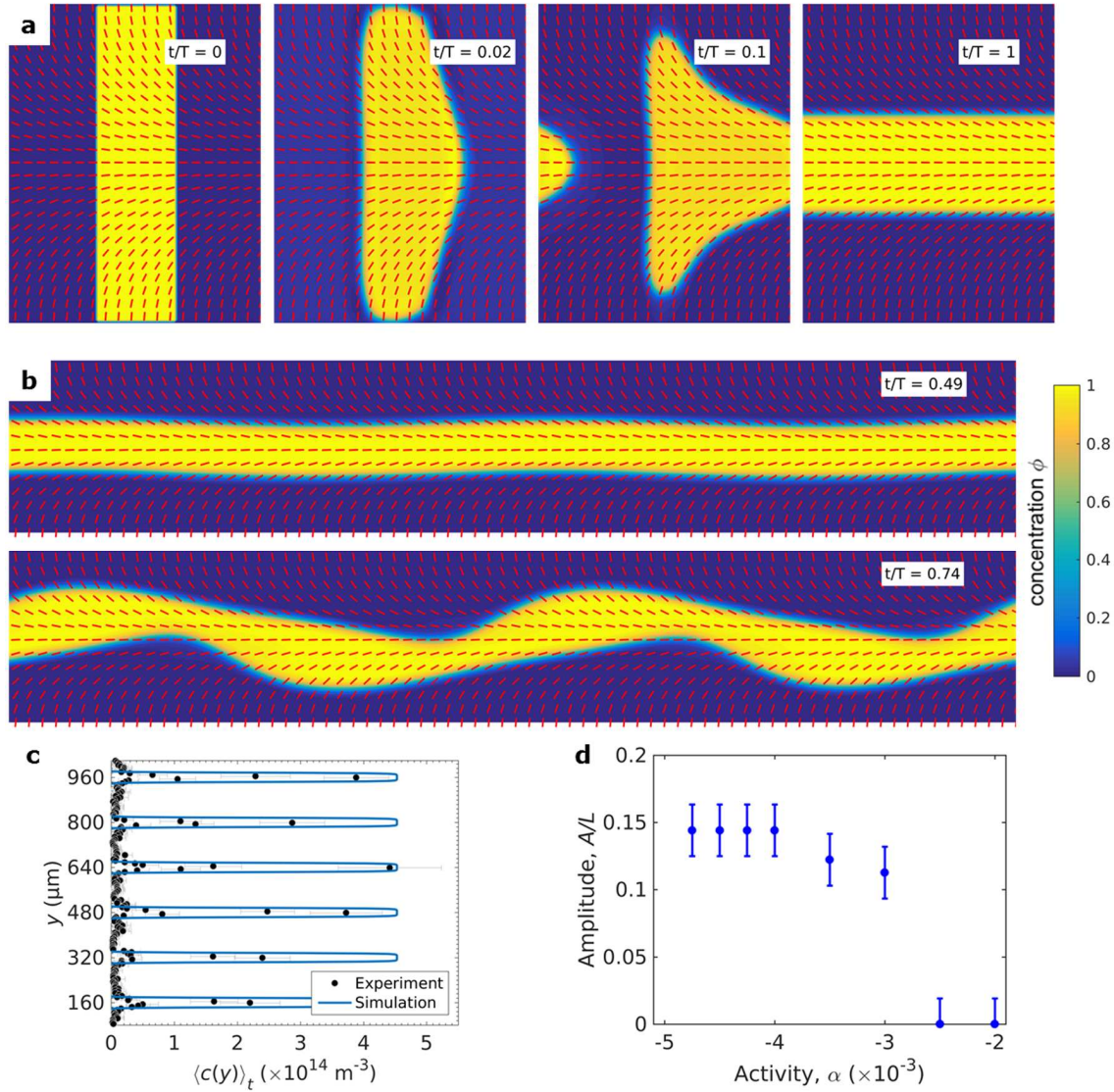

**Fig. 4.** Simulation results for the extensile living nematic ($\alpha < 0$) based on the two-phase model. **a**, The active nematic concentrates into splay region and moves from left to right. The simulation is initialized with a blob of active phase ($\phi = 1$). The red solid lines show the directors of the background passive liquid crystal. **b**, For higher activity, the jet starts to undulate. **c**, Comparison of the experimental (dots; data for the dispersion with $\langle c \rangle \approx 0.3 \times 10^{14}$ m$^{-3}$) and simulated (line) concentration profiles for extensile active particles focusing into unidirectional jets within splay regions. The theoretical curves are normalized by the maximum concentration of swimmers in the splay regions. **d**, The steady-state amplitude of the undulation as a function of the dimensionless activity. The amplitude $A$ is defined as the maximum height of the centre of the stripe.

Consistent with the experiments, the unidirectional jet starts to undulate once the activity is sufficient (Fig. 4b and Supplementary Video 7). The underlying director pattern stabilizes the undulation with a well-defined final amplitude that depends on the activity (Fig. 4d). In the simulations it is possible to further increase the activity. At high activities, the undulations continue to grow beyond the bend region in the background pattern, giving rise to an active turbulent state.

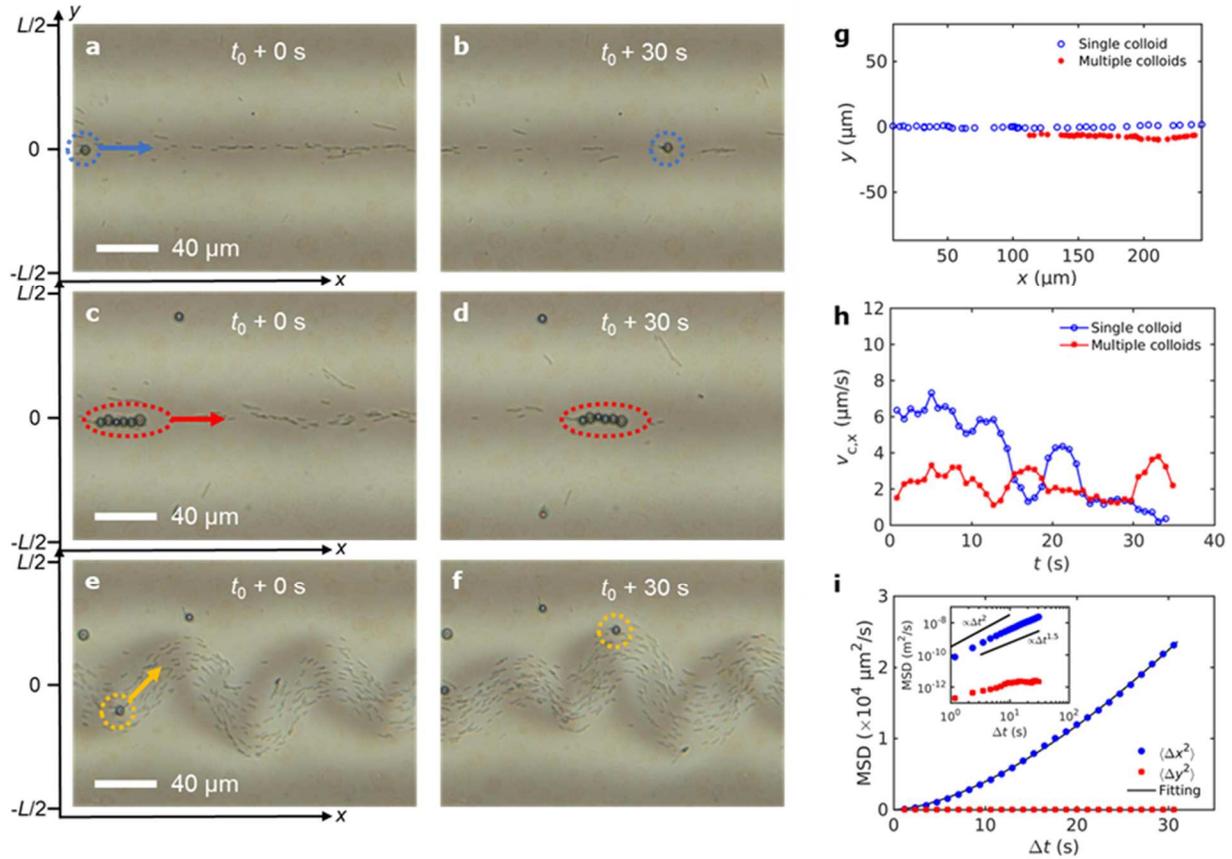

**Fig. 5**. Cargo transport by the rectilinear and undulating bacterial jets. **a,b,** Optical microscope images taken 30 s apart of a single 5 μm diameter glass sphere and **c,d,** aggregate consisting of 6 glass spheres located in the splay region. The particles are relocated from left to right by the incoming unidirectional bacterial flow. **e,f,** Colloidal particle carried by the undulating jet of bacteria. **g,** The trajectories and **h,** instantaneous horizontal component of the velocities of colloidal particles carried by the rectilinear jet. **i,** Mean squared displacement (MSD) of the single colloid being transported by the jet in **a,b**; dashed line is the least square power fit to MSD along the *x*-axis.

The condensed bacterial jets can serve as effective micro-cargo transporters. As an example, Fig. 5 and Supplementary Videos 8-10 show transport of 5 μm diameter glass colloidal spheres by rectilinear jets (Fig. 5a-d), with a velocity in a range 1-7 μm/s (Fig. 5g,h) and by undulating jets (Fig. 5e,f). The splay-bend patterning in Fig. 5a-d allows one to transport micro-cargo along a well-defined $y$-coordinate in the $xy$-plane of the cell. This is an advantage over the unipolar transport reported earlier (*6, 27, 28*) that does not select a particular $y$-coordinate. Considering the momentum $p = mv_c$ of cargo, where $m$ is its mass and $v_c$ is its speed, the momentum transferred by the bacterial jets to a single colloid is $p_{single} \approx 50 \times 10^{-20}$ kg m s$^{-1}$ and to a chain of six colloids is $p_{aggreg} \approx 200 \times 10^{-20}$ kg m s$^{-1}$ (Fig. 5a-d). These momenta are a few orders of magnitude higher than the momentum $(0.001-2) \times 10^{-20}$ kg m s$^{-1}$ imparted by single bacteria onto smaller particles (*27, 28*). The calculated mean square displacement (MSD) of the single colloid shows the strongly ballistic transport along the jet of bacteria and confined diffusion in the perpendicular direction. Interpolation of the MSD along the $x$-axis with $\langle \Delta x^2 \rangle = 2D_{eff} \Delta t^\gamma$, where $D_{eff}$ is the anomalous diffusion coefficient, $\gamma$ is the anomalous diffusion exponent, and $\Delta t$ is the time-lag, yields $\gamma = 1.6 \pm 0.1$, indicating superdiffusive behavior of the sphere. The anomalous diffusion persists at the time scales of the experiment (30 s) without any signs of reducing to normal diffusion. Thus, the patterned director field extends the superdiffusion to the time scales much higher than the time scales limit of a few seconds observed for anomalous diffusion in an isotropic bacterial bath (*29, 30*).

In conclusion, we demonstrate the different roles of splay and bend in the patterned nematic director that controls collective motion of swimming bacteria. The bacteria engage in threshold-less unidirectional collective motion along the splay bands. The focusing is explained by the aligning action of the underlying director onto the orientation of bacteria. The pre-imposed pattern makes the focusing persistent as the bacterial concentration increases. At very high concentrations, rectilinear jets start to undulate because of the hydrodynamic bend instability of interactive extensile active units. However, undulations are stabilized by the background director pattern and do not grow beyond the bend regions. The underlying nematic pattern thus prevents the instabilities of collective dynamics that are an intrinsic property of aligned polar self-propelled particles swimming in an isotropic environment (*11, 23, 24*). Both unidirectional jet and undulating streams are capable of transporting colloidal particles, indicating possible future applications of this set-up in microfluidics and micro-cargo transport.

**Materials and methods**

*Bacteria dispersion preparation*

We used solution of disodium chromoglycate (DSCG) (purchased from *Alfa Aesar*), in aqueous solution of terrific broth (TB) medium (*Sigma Aldrich*) as a nematic host. TB solution serves as a growing and motility medium for *Bacillus subtilis* (strain 1085). The bacteria were inoculated from the lysogeny broth (LB) agar plate (*Teknova*) into the tube with 10 mL of TB and then grown in a sealed tube to adapt the bacteria to oxygen starvation at 35°C inside the shaking incubator. The concentration of the bacteria was monitored with the custom-made optical density meter (based on Raspberry Pi 3 microcomputer) and the tube was extracted from the incubator before saturation of the bacterial concentration at $10^{13}$ m$^{-3}$. 0.1 mL of the liquid medium with bacteria was centrifuged at 8000 rpm for 2.5 min and the medium was extracted upon which 0.1 mL of 14 wt% solution of DSCG in TB was added to the bacteria and carefully stirred with the pipet. The dispersion of *B. subtilis* in DSCG was used before the bacteria lost their swimming ability which is typically within 1-2 hours in a sealed cell without an oxygen supply. To obtain a concentration of bacteria above $10^{13}$ m$^{-3}$, we centrifuged 0.5 ml TB medium containing a high concentration of bacteria, extracted the TB and added 0.1 ml of DSCG to the tube and stirred it with the pipet. This bacterial suspension was injected into the cell and observed under the microscope.

*Sample preparation*

For the photopatterned cells we cleaned soda-lime glass sheet about 1 mm in thickness in the ultrasonic bath with detergent at 60°C, rinsed it with isopropanol and dried it in the oven at 90°C for 10 min. The glass was also treated with UV-ozone in a sealed chamber for 10 min and after that it was spincoated at 3000 rpm for 30 seconds with the azo-dye 0.5 wt% filtered solution of Brilliant Yellow (*Sigma Aldrich*) in dimethylformamide. The solvent was evaporated by annealing the glass substrates at 100°C for 30 min in the oven. The glass substrates with photoalignment layer were illuminated with a metal-halide X-Cite 120 lamp through the plasmonic photomask with a polarization map created by array of nanoslits. The substrates were cleaned with a compressed nitrogen gun, spincoated with toluene solution of 6.7 wt% reactive mesogen RM-257 (*Wilshire Tech.*) and 0.35 wt% photoinitiator Irgacure-651 (*Ciba*), illuminated with a 365 nm ultra violet lamp (UVL-56, 6 W) for 30 min to completely polymerize reactive mesogen(*7, 19*).

Colloidal glass spheres used as cargo were purchased from *Bangs Laboratories, Inc.*; their mass density is $\approx 2.2 \times 10^3$ kg m$^{-3}$. The cells were sealed by the epoxy resin *Devcon*.

***Data acquisition and analysis***

We used a Nikon TE-2000 inverted microscope with high resolution camera Emergent HR-20000C and 20× magnification objective. The video sequences were recorded at 2 and 5 Hz rate. The sequences of images were analyzed by ImageJ2-Fiji(*31, 32*), particle image velocimetry MATLAB package(*33*) and custom written MATLAB code. To determine the local concentration dependence $c(y)$, we count the bacteria from the microscopic images with enhanced contrast. The bacteria appear as higher intensity objects on a lower background intensity of the transparent liquid crystal field. We find the average image over all frames and subtract it from each frame of the video. With this the background liquid crystal intensity becomes even more suppressed and we segregate the individual bacteria using intensity thresholding technique. We then divide the number of bacteria in a narrow stripe of width $\Delta y$ by the stripe volume $\Delta V = \Delta y W d$, where $W = 736 \, \mu m$ is the width of the image along the $x$-axis. The concentration of bacteria in an undulating jet was calculated as the total number of bacteria $n$ in the jet divided by the volume of the jet, which is proportional to its contour length $l_c$ and width $w$: $c = n / (w l_c d)$ (Fig. 3d and Supplementary Fig. 2b).

**References**

1.  M. C. Marchetti *et al.*, Hydrodynamics of soft active matter. *Reviews of Modern Physics* **85**, 1143-1189 (2013).

2.  S. Zhou, A. Sokolov, O. D. Lavrentovich, I. S. Aranson, Living liquid crystals. *Proceedings of the National Academy of Sciences* **111**, 1265-1270 (2014).

3.  I. I. Smalyukh, J. Butler, J. D. Shrout, M. R. Parsek, G. C. L. Wong, Elasticity-mediated nematiclike bacterial organization in model extracellular DNA matrix. *Phys Rev E* **78**, 030701 (2008).

4.  A. Kumar, T. Galstian, S. K. Pattanayek, S. Rainville, The motility of bacteria in an anisotropic liquid environment. *Molecular Crystals and Liquid Crystals* **574**, 33-39 (2013).

5.  P. C. Mushenheim, R. R. Trivedi, H. H. Tuson, D. B. Weibel, N. L. Abbott, Dynamic self-assembly of motile bacteria in liquid crystals. *Soft Matter* **10**, 88-95 (2014).

6.  P. C. Mushenheim *et al.*, Effects of confinement, surface-induced orientations and strain on dynamical behaviors of bacteria in thin liquid crystalline films. *Soft Matter* **11**, 6821-6831 (2015).

7.  C. Peng, T. Turiv, Y. Guo, Q. H. Wei, O. D. Lavrentovich, Command of active matter by topological defects and patterns. *Science* **354**, 882-885 (2016).

8.  M. M. Genkin, A. Sokolov, O. D. Lavrentovich, I. S. Aranson, Topological defects in a living nematic ensnare swimming bacteria. *Phys Review X* **7**, 1-14 (2017).

9.  M. M. Genkin, A. Sokolov, I. S. Aranson, Spontaneous topological charging of tactoids in a living nematic. *New Journal of Physics* **20**, 043027 (2018).

10. A. Sokolov, I. S. Aranson, Physical properties of collective motion in suspensions of bacteria. *Phys Rev Letters* **109**, 248109(248105) (2012).

11. R. Aditi Simha, S. Ramaswamy, Hydrodynamic fluctuations and instabilities in ordered suspensions of self-propelled particles. *Phys Rev Letters* **89**, 058101 (2002).

12. R. Green, J. Toner, V. Vitelli, Geometry of thresholdless active flow in nematic microfluidics. *Phys Review Fluids* **2**, 104201 (2017).

13. T. Sanchez, D. T. N. Chen, S. J. Decamp, M. Heymann, Z. Dogic, Spontaneous motion in hierarchically assembled active matter. *Nature* **491**, 431-434 (2012).

14. S. Ramaswamy, The Mechanics and Statistics of Active Matter. *Annual Review of Condensed Matter Physics* **1**, 323-345 (2010).

15. F. C. Keber *et al.*, Topology and dynamics of active nematic vesicles. *Science* **345**, 1135-1139 (2014).

16. P. Guillamat, J. Ignés-Mullol, F. Sagués, Control of active liquid crystals with a magnetic field. *Proceedings of the National Academy of Sciences* **113**, 5498-5502 (2016).

17. P. Guillamat *et al.*, Active nematic emulsions. *Science Advances* **4**, 1-11 (2018).

18. A. Maitra *et al.*, A nonequilibrium force can stabilize 2D active nematics. *Proceedings of the National Academy of Sciences* **115**, 6934-6939 (2018).

19. C. Peng *et al.*, Patterning of Lyotropic Chromonic Liquid Crystals by Photoalignment with Photonic Metamasks. *Advanced Materials* **29**, 1606112 (2017).

20. G. E. Volovik, O. D. Lavrentovich, Topological dynamics of defects: boojums in nematic drops. *Sov. Phys. JETP* **58**, 1159-1166 (1983).

21. S. Zhou *et al.*, Dynamic states of swimming bacteria in a nematic liquid crystal cell with homeotropic alignment. *New Journal of Physics* **19**, 055006 (2017).

22. R. A. Simha, S. Ramaswamy, Statistical hydrodynamics of ordered suspensions of self-propelled particles: Waves, giant number fluctuations and instabilities. *Physica A: Statistical Mechanics and its Applications* **306**, 262-269 (2002).

23. A. Be'er, G. Ariel, A statistical physics view of swarming bacteria. *Movement Ecology* **7**, 9 (2019).

24. D. Saintillan, M. J. Shelley, Instabilities and pattern formation in active particle suspensions: Kinetic theory and continuum simulations. *Phys. Rev. Lett.* **100**, 1-4 (2008).

25. J. W. Cahn, Free energy of a nonuniform system. II. Thermodynamic basis. *The Journal of Chemical Physics* **30**, 1121-1124 (1959).

26. M. L. Blow, S. P. Thampi, J. M. Yeomans, Biphasic, lyotropic, active nematics. *Phys Rev Letters* **113**, 248303 (2014).

27. A. Sokolov, S. Zhou, O. D. Lavrentovich, I. S. Aranson, Individual behavior and pairwise interactions between microswimmers in anisotropic liquid. *Phys Rev E* **91**, 013009 (2015).

28. R. R. Trivedi, R. Maeda, N. L. Abbott, S. E. Spagnolie, D. B. Weibel, Bacterial transport of colloids in liquid crystalline environments. *Soft Matter* **11**, 8404-8408 (2015).

29. X.-L. Wu, A. Libchaber, Particle Diffusion in a Quasi-Two-Dimensional Bacterial Bath. *Phys Rev Letters* **27**, 3017-3020 (2000).

30. C. Valeriani, M. Li, J. Novosel, J. Arlt, D. Marenduzzo, Colloids in a bacterial bath: Simulations and experiments. *Soft Matter* **7**, 5228-5238 (2011).

31. J. Schindelin *et al.*, Fiji: An open-source platform for biological-image analysis. *Nature Methods* **9**, 676-682 (2012).

32. J. Y. Tinevez *et al.*, TrackMate: An open and extensible platform for single-particle tracking. *Methods* **115**, 80-90 (2017).

33. W. Thielicke, E. J. Stamhuis, PIVlab – Towards User-friendly, Affordable and Accurate Digital Particle Image Velocimetry in MATLAB. *Journal of Open Research Software* **2**, e30 (2014).

**Acknowledgement**

This work was supported by NSF grants DMS-1729509 and CMMI-1436565. Research of I.S.A. was supported by the NSF grant PHY-1707900. K.T. was funded by the European Union's Horizon 2020 research and innovation programme under the Marie Sklodowska-Curie grant agreement No 722402. A. D. was supported by a Royal Commission for the Exhibition of 1851 Research Fellowship. We would like to acknowledge valuable discussions with J. Toner, V. Vitelli, and R. Green. We also thank S. Shiyanovskii and N. Aryasova who provided the Mathematica code for PolScope images analysis.

# SUPPLEMENTARY INFORMATION

**Supplementary text**

***Realignment of a bacterium away from the director of the surrounding nematic.*** A bacterium such as *B. subtilis* can develop a realigning torque $\tau_{\text{bact}} \approx \beta f l_{\text{D}}$, where $\beta$ is the angle between the bacterial body axis and the local $\hat{\mathbf{n}}$, $f$ is the thrust force, and $l_{\text{D}}$ is the length of the force dipole associated with the force-free mode of locomotion. In the order of magnitude (*21*), $\tau_{\text{bact}} / \beta \approx (1-10) \times 10^{-18} \text{ N m}$. The stabilizing torque of the surrounding nematic on a bacterium attempting a realignment comes from the elastic distortions and from finite strength of the surface anchoring of the director at the bacterial body. The elastic torque, calculated by Smith and Denniston (*34*) for a rod of length $l$ and radius $r$ with an infinitely strong anchoring at its surface, realigned by an angle $\beta$ from $\hat{\mathbf{n}}$, is $\tau_{\text{el}} \approx \beta CKl$, where $C \approx 4\pi / \ln(2l/r)$. For the typical $K = 10$ pN (*35*), $l = 8$ $\mu$m, $r = 0.4$ $\mu$m, one finds $\tau_{\text{el}} / \beta \approx 3 \times 10^{-16} \text{ N m}$, a value much higher than $\tau_{\text{bact}} / \beta$ and thus prohibitively high for the bacterium to realign. Of course, the assumption of infinite surface anchoring must be mitigated and the surface energy penalty for the bacterial realignment with respect to $\hat{\mathbf{n}}$ should be finite, $\frac{1}{2} W \beta^2$, where $W$ is the surface anchoring strength (*21*). The value of $W$ is expected to be on the order of $10^{-6}$ J/m$^2$ or even somewhat less (*36-38*). The corresponding stabilizing anchoring torque is then reduced to $\tau_{\text{surf}} / \beta \approx 2\pi r l W \approx (10-20) \times 10^{-18} \text{ N m}$, comparable but still typically larger than the bacterial realigning torque $\tau_{\text{bact}} / \beta \approx (1-10) \times 10^{-18} \text{ N m}$. We thus conclude that the preferred direction of swimming is parallel to the local $\hat{\mathbf{n}}$.

***Reversal time of the bacteria swimming direction.*** The probability for the bacteria to reverse the direction of motion was calculated as the ratio of the number of reversal events to the total number of bacteria swimming through the patterned director field over the period of $\Delta t = 60$ s (Supplementary Video 1). The reversal probability $p_0 \sim 5 \times 10^{-3}$ was calculated for three intervals, each of time interval $\Delta t$ and averaged. The experimental reversal time of the bacteria swimming direction was then calculated as $\tau = \Delta t / p_0 \sim 1.2 \times 10^4$ s.

***Bending instability in the cell with uniform alignment.*** The uniformly aligned cell with $\hat{\mathbf{n}}$ being oriented along horizontal $x$-axis was used to measure the critical concentration above which the bending instability occurs (Supplementary Fig. 1a and Supplementary Video 2). The thickness of the cell and bacteria activity was chosen to be the same as in patterned cells. We focused on the region where the local bacterial concentration changes along the vertical $y$-axis and calculated the concentration of bacteria within a thin band of height $\Delta y = 5$ $\mu$m. From the time averaged concentration distribution of bacteria along the $y$-axis (Supplementary Fig. 1b,c) we find the critical concentration after which the bending instability sets in, in the horizontal band between 100 and 250 $\mu$m, to be $c_0 \approx 0.9 \times 10^{14}$ m$^{-3}$.

***Advection-diffusion analytical model for living nematic.*** The constants $C^\pm$ in equation (5) can be determined from the following consideration. According to the definition, a population $c^+$ swims parallel to $\hat{\mathbf{n}}$, and $c^-$ swims antiparallel. Thus, the direction of motion of $c^+$ and $c^-$ alternates between the neighboring bands. Because of these alternations, $c^+$ in one band travels in the same direction as $c^-$ in the neighboring band (e.g. from left to right). This makes it possible to construct a nematic $L$-periodic solution rather than a $2L$-periodic solution corresponding to the periodicity of the vector field $\hat{\mathbf{n}}$. Assuming $c^+(y) = c^-(y+L)$, we find $C^+ = C^- = C_0/2$.

***Advection-diffusion computational model for living nematic.*** To investigate the experimentally observed undulations we used a computational model that combines the Edwards-Beris description of nematic LC with advection-diffusion bacterial dynamics, see (*8, 9*). The resulting system of coupled partial differential equations was numerically integrated on Graphical Processor Units. In this work we only made a few minor modifications of the model described in (*8*):

- To incorporate the patterned director field, we made the external aligning term $\mathbf{F}_{ext}$ in the equation for the nematic director spatially-dependent (*8*), so that the aligning direction is given by equation (1) in the main text. In our framework the nematic director is allowed to deviate from the patterned direction, while the degree of deviation is controlled by the amplitude of the $\mathbf{F}_{ext}$ term (we set $|\mathbf{F}_{ext}| = 0.1$ in this work).

- The parameter values of our computational model can be found in Ref. (*8*). In this work we varied the average bacterial concentration $C_0$ and bacterial diffusion coefficient $D_c$ to match the experimental data. In addition, we reduced the depth-averaging coefficient magnitude $\zeta = 0.07\eta/h^2$ (compare to (*8*)), where $\eta$ is the effective nematic viscosity, $h$ is the cell thickness. In the DSCG, the bending viscosity is much smaller than twist and splay viscosities (*35*), which reduces the effective viscosity in the expression for depth-averaging term. The value 0.07 was chosen by fitting our model to the experimental data (*2*). All other parameters were the same as in (*8*).

- In our two-populational bacterial model ($c^+$ swims parallel and $c^-$ swims antiparallel to the nematic) there was an issue near the places with vertical nematic orientation ($\theta = \pi/2$), where the vector nematic angle is discontinuous (although the nematic director is continuous). To address this issue, we relabeled $c^\pm$ concentrations in the detected problematic places in (*8*). In this work the nematic direction is approximately given by equation (1), and it is possible to introduce a continuous nematic vector orientation (Fig. 1b). This is achieved by changing $\theta \in (-\pi/2; \pi/2]$ to $\theta \in (3\pi/2; 2\pi]$ in the adjacent regions, separated by bending bands. With the continuous nematic vector angle there is no need for relabeling $c^\pm$.

- In order to examine how the concentration distribution $c(y)$ is affected by the finite reversal time $\tau$ of swimming direction, we followed Ref. (*8*) and performed numerical integration of equation (3) and coupled order parameter dynamics equations, where $\tau$ was set to 60 s and 600 s (Fig. 2d and Supplement Video 4). The only additional parameter is the depth-averaged coefficient $\zeta$, which was reduced by a factor of 200 compared to the one in (8). The reason behind this is the drastic difference between the DSCG viscosities (measured in Zhou et. al. (35)), which can make effective isotropic viscosity smaller.

- The existence of symmetric nonzero vertical components of the velocity in the simulations, clearly visible in Fig. 2h, comes from the fact that the bacteria from different concentration subsets would converge into splay regions along the positive or negative $y$ direction, thus, acquiring a small positive or negative $v_y$ component. However, the simulations yield no net flows along the $y$-axis, in agreement with the experiment.

***Computational model for two-phase continuum description.*** We use numerical simulations of active nemato-hydrodynamics to solve for the total density $\rho$ and the velocity $u$. The orientational order of the bacteria is described by the nematic tensor $\mathbf{Q} = \frac{3q}{2}\left(\hat{\mathbf{n}}\hat{\mathbf{n}} - \mathbf{I}\frac{1}{3}\right)$, where $q$ denotes the magnitude of the orientational order and $\hat{\mathbf{n}}$ is the director (39). This nematic tensor obeys (40)

$$\left(\partial_t + \mathbf{u}\cdot\nabla\right)\mathbf{Q} - \mathbf{S} = \Gamma_{\mathbf{Q}}\mathbf{H}, \tag{7}$$

where $\Gamma_{\mathbf{Q}}$ is the rotational diffusivity and $\mathbf{S}$ is the co-rotational advection term that accounts for the impact of the strain rate $\mathbf{E} = \frac{1}{2}\left(\nabla\mathbf{u}^T + \nabla\mathbf{u}\right)$ and vorticity $\mathbf{\Omega} = \frac{1}{2}\left(\nabla\mathbf{u}^T - \nabla\mathbf{u}\right)$ on the director field. This co-rotational advection has the form

$$\mathbf{S} = \left(\lambda\mathbf{E} + \mathbf{\Omega}\right)\cdot\left(\mathbf{Q} + \mathbf{I}\frac{1}{3}\right) + \left(\mathbf{Q} + \mathbf{I}\frac{1}{3}\right)\cdot\left(\lambda\mathbf{E} - \mathbf{\Omega}\right) - 2\lambda\left(\mathbf{Q} + \mathbf{I}\frac{1}{3}\right)(\mathbf{Q}:\nabla\mathbf{u}), \tag{8}$$

where the alignment parameter $\lambda$ is related to the shape of the active particles with $\lambda > 0$ for rod-like particles. The relaxation of the orientational order is controlled by the molecular field

$$\mathbf{H} = -\left(\frac{\delta\mathcal{F}}{\delta\mathbf{Q}} - \frac{1}{3}\mathbf{I}Tr\left(\frac{\delta\mathcal{F}}{\delta\mathbf{Q}}\right)\right), \tag{9}$$

which models the relaxation towards the minimum of a free energy $\mathcal{F}$.

We introduce an order parameter $\phi$ to distinguish between the regions where the bacteria are absent $\phi \approx 0$ and present $\phi \approx 1$. This approach has been used in the past to model growing cell colonies (41). The assumption here is that we can approximate the concentration of bacteria as uniform and constant in the concentrated bacterial regions. The order parameter $\phi$ follows a Cahn-Hillard equation $\left(\partial_t\phi + \nabla\cdot(\mathbf{u}\phi)\right) = \Gamma_\phi\mu$, where $\Gamma_\phi$ is the mobility of the concentration and $\mu = \frac{\delta\mathcal{F}}{\delta\phi} - \nabla\cdot\left(\frac{\delta\mathcal{F}}{\delta\nabla\phi}\right)$ is the chemical potential which is the functional derivative of the free energy with respect to $\phi$.

The total free energy density $\mathcal{F} = \mathcal{F}_\mathbf{Q} + \mathcal{F}_\phi + \mathcal{F}_{LC}$ contains three components. The first component

$$\mathcal{F}_\mathbf{Q} = \frac{1}{2} A(\phi - 0.5) Tr(\mathbf{Q}^2) + C Tr(\mathbf{Q}^2)^2 + \frac{1}{2} K(\nabla \mathbf{Q})^2 \qquad (10)$$

is a modified Landau-de Gennes bulk free energy density that couples the $\phi$ and the nematic order parameter. This ensures that we have nematic ordering in the high $\phi$ regions. This free energy term also contains the elastic free energy density due to spatial inhomogeneities in the nematic ordering. Here $A$, $C$, and $K$ are material constants. The second free energy term

$$\mathcal{F}_\phi = \frac{1}{2} A_\phi \phi^2 (1 - \phi)^2 + \frac{1}{2} K_\phi (\nabla \phi)^2 \qquad (11)$$

is a Ginzburg-Landau free energy to allow for phase ordering and surface tension between the active nematic and the surrounding fluid. Here $A_\phi$ and $K_\phi$ are again material constants. Lastly, we include the term

$$\mathcal{F}_{LC} = \frac{1}{2} W_c Tr(\mathbf{Q} - \mathbf{Q}_S)^2 \qquad (12)$$

to model the effect of the underlying liquid crystal which has a director field $\mathbf{Q}_S$. We contain all bacteria-liquid crystal interactions in this Rapini-Papoular like free energy penalty, which penalizes orientations of the bacteria that are not in alignment with the underlying liquid crystal. $W_c$ takes the form of an effective anchoring strength which is taken high in our simulations as previous work shows that bacteria quickly relax to the underlying liquid crystal orientation (*8*). Since the experiments are performed in a thin-film geometry with the scale of the emergent pattern much larger than the cell thickness, we use a quasi-two-dimensional approximation. We also argue that the anchoring of the underlying liquid crystal to the container walls is sufficiently strong compared to the bacteria-liquid crystal interaction that all perturbations in the liquid crystal are negligible and that $\mathbf{Q}_S$ is constant in time. This approximation is sufficient to reproduce the experimental observations.

The local density and velocity field obey the incompressible Navier-Stokes equations

$$\nabla \cdot \mathbf{u} = 0, \qquad (13)$$

$$\rho(\partial_t + \mathbf{u} \cdot \nabla)\mathbf{u} = \nabla \cdot \mathbf{\Pi}, \qquad (14)$$

where $\mathbf{\Pi}$ is the generalized stress tensor that includes both nematic, active and capillary contributions, in addition to the viscous stress $\mathbf{\Pi}^{\text{visc}} = 2\mu \mathbf{E}$. The stress due to elastic contributions arising from nematic ordering of the bacteria is

$$\mathbf{\Pi}^{\text{el}} = -P\mathbf{I} + 2\lambda\left(\mathbf{Q} + \mathbf{I}\frac{1}{3}\right)(\mathbf{Q} : \mathbf{H}) - \lambda \mathbf{H} \cdot \left(\mathbf{Q} + \mathbf{I}\frac{1}{3}\right) - \lambda\left(\mathbf{Q} + \mathbf{I}\frac{1}{3}\right) \cdot \mathbf{H} - \nabla \mathbf{Q} : \frac{\delta \mathcal{F}}{\delta \nabla \mathbf{Q}} + \mathbf{H} \cdot \mathbf{Q} - \mathbf{Q} \cdot \mathbf{H}, \quad (15)$$

which includes the bulk pressure $P$ (*40*).

We also include capillary stresses $\mathbf{\Pi}^{\text{cap}} = (\mathcal{F} - \mu\phi)\mathbf{I} - \nabla\phi \frac{\delta \mathcal{F}}{\delta \nabla \phi}$ as we work with a two-component system. Lastly, the active stress accounts for changes in the flow field caused by continual energy injection at the microscopic scale. Activity generates flows for nonzero gradients of $\mathbf{Q}$ and takes the form $\mathbf{\Pi}^{\text{active}} = \alpha \mathbf{Q}$ (11). The activity parameter $\alpha$ determines the strength of the active flows with negative and positive values denoting extensile and contractile fluids, respectively. It represents the local average energy injection per unit area of the bacteria. This means that it should scale with the concentration of bacteria. In our simulations, we take this parameter to be constant in space and time, as we previously approximated the bacterial density to be constant in the high concentration regimes. The equations of active nemato-hydrodynamics are solved using a hybrid lattice Boltzmann scheme.

In the simulations, the free energy parameters are $A = -0.1$, $C = 1.2$, $K = 0.025$, $A_\phi = 0.05$, $K_\phi = 0.025$, and $W_c = 0.008$. The tumbling parameter $\lambda$ is set to 0.3 and the viscosity $\mu$ is $0.667$. Unless otherwise states, the activity $\alpha$ is $-0.003$. All parameters are in lattice Boltzmann units, where the density, lattice size and time step are taken as unity.

**References**

1. S. Zhou et al., Dynamic states of swimming bacteria in a nematic liquid crystal cell with homeotropic alignment. New Journal of Physics **19**, 055006 (2017).
2. C. J. Smith, C. Denniston, Elastic response of a nematic liquid crystal to an immersed nanowire. Journal of Applied Physics **101**, 014305 (2007).
3. S. Zhou et al., Elasticity, viscosity, and orientational fluctuations of a lyotropic chromonic nematic liquid crystal disodium cromoglycate. Soft Matter **10**, 6571-6581 (2014).
4. H. S. Park, S. W. Kang, L. Tortora, S. Kumar, O. D. Lavrentovich, Condensation of self-assembled lyotropic chromonic liquid crystal sunset yellow in aqueous solutions crowded with polyethylene glycol and doped with salt. Langmuir **27**, 4164-4175 (2011).
5. V. G. Nazarenko et al., Surface alignment and anchoring transitions in nematic lyotropic chromonic liquid crystal. Phys Rev Letters **105**, 017801 (2010).
6. P. J. Collings, P. van der Asdonk, A. Martinez, L. Tortora, P. H. J. Kouwer, Anchoring strength measurements of a lyotropic chromonic liquid crystal on rubbed polyimide surfaces. Liquid Crystals **44**, 1165-1172 (2017).
7. M. M. Genkin, A. Sokolov, O. D. Lavrentovich, I. S. Aranson, Topological defects in a living nematic ensnare swimming bacteria. Phys Review X **7**, 1-14 (2017).
8. M. M. Genkin, A. Sokolov, I. S. Aranson, Spontaneous topological charging of tactoids in a living nematic. New Journal of Physics **20**, 043027 (2018).
9. S. Zhou, A. Sokolov, O. D. Lavrentovich, I. S. Aranson, Living liquid crystals. Proceedings of the National Academy of Sciences **111**, 1265-1270 (2014).
10. P. G. d. Gennes, J. Prost, The physics of liquid crystals. 597 (1993).
11. A. N. Beris, B. J. Edwards, Thermodynamics of Flowing Systems. 704 (1994).
12. A. Doostmohammadi, S. P. Thampi, J. M. Yeomans, Defect-Mediated Morphologies in Growing Cell Colonies. Phys Rev Letters **117**, 048102 (2016).
13. R. Aditi Simha, S. Ramaswamy, Hydrodynamic fluctuations and instabilities in ordered suspensions of self-propelled particles. Phys Rev Letters **89**, 058101 (2002).

**Supplementary figures**

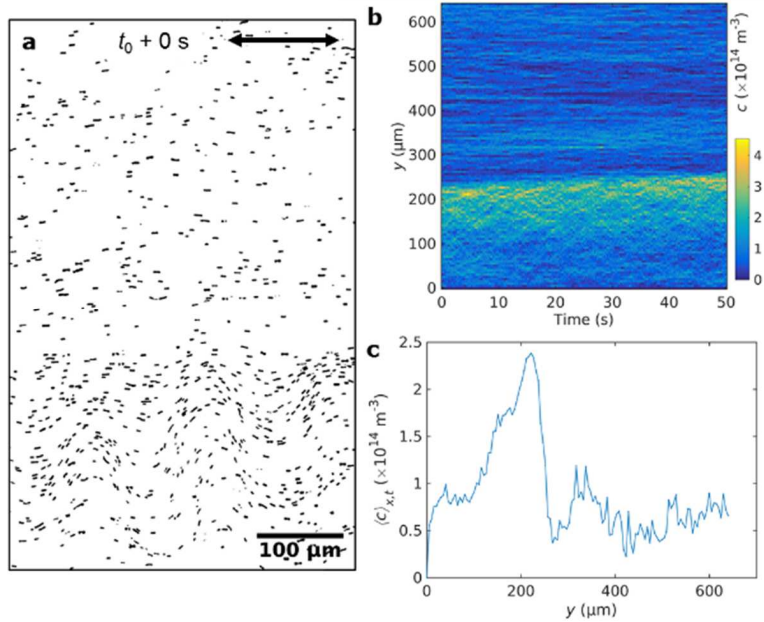

**Fig. S1**. Bacteria in uniform nematic (cell thickness 20 μm). **a**, Enhanced contrast microscopic image of bacteria swimming through uniformly aligned nematic cell (orientation axis is given by double-headed arrow) with a concentration gradient. Bacteria exhibit both uniform swimming (top part) and bending instability (bottom part). **b**, Time evolution of the concentration of bacteria along $y$-axis with the length step $\Delta y = 5$ μm and time step 0.5 s. **c**, Averaged concentration over time (50 s) and width of the image. Bacteria concentration at which the undulation occurs, is $c_\mathrm{u}^\mathrm{uniform} \approx 0.9 \times 10^{14}$ m$^{-3}$, much lower than the threshold concentration for the undulation of condensed jets in patterned splay-bend cells, $c_\mathrm{u}^\mathrm{s-b} \approx 11.8 \times 10^{14}$ m$^{-3}$.

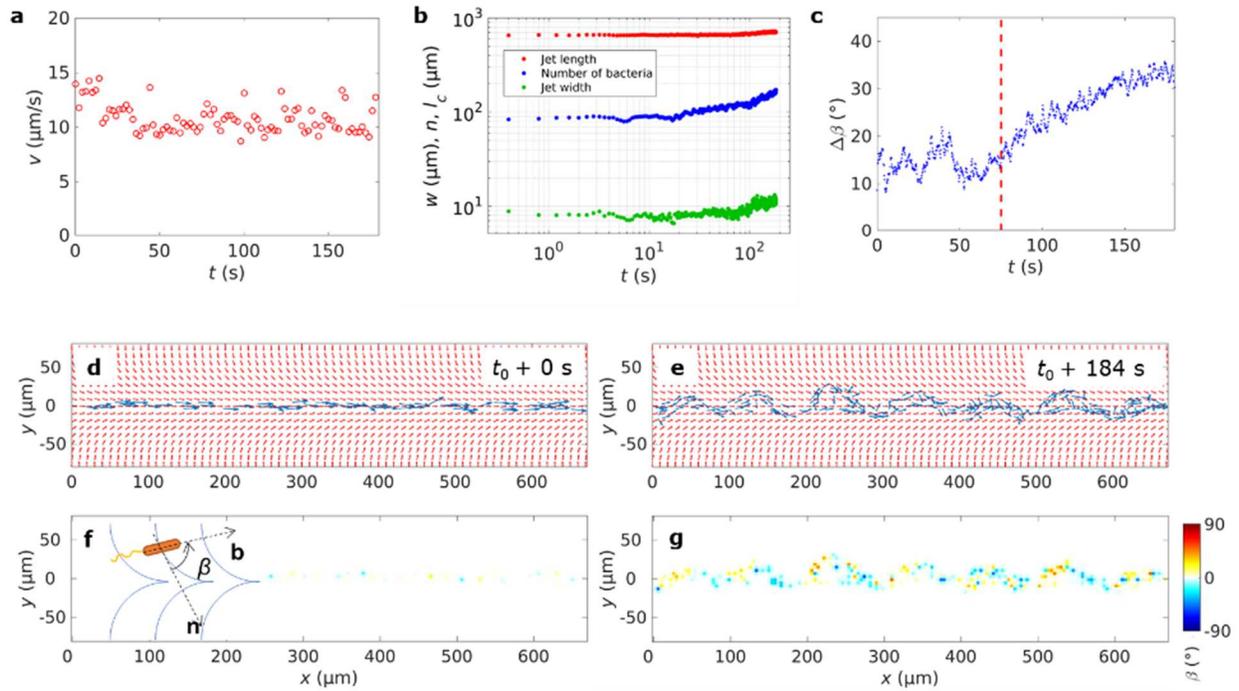

**Fig. S2.** Undulation of a bacterial jet in a highly concentrated dispersion $\langle c \rangle \approx 1.5 \times 10^{14}$ m$^{-3}$ in a patterned cell of thickness 20 μm. **a**, Average bacteria speed in the developing undulation. **b**, Time evolution of the jet contour length, width and number of bacteria. **c**, Standard deviation of angle $\beta$ between the bacteria body axes and the director imposed at the substrates, as a function of time. **d**, Pre-imposed director (red arrows) and bacterial axes (blue arrows) in a uniform jet. **e**, The same for an undulating jet. **f** Definition of the angle $\beta$ between the bacteria and the pre-imposed director. **g**, Map of the angle $\beta$ for the undulating jet shown in part **e**.

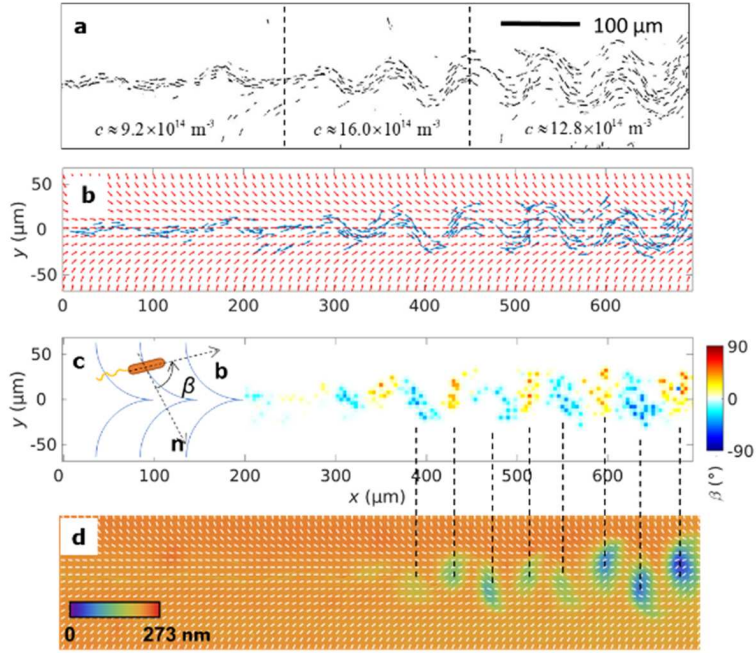

**Fig. S3.** Undulation instability of a polar jet in the highly concentrated dispersion, $\langle c \rangle \approx 1.5 \times 10^{14}$ m$^{-3}$, imaged by PolScope in a 10 µm thick cell. **a,** Enhanced contrast microscopic snapshots of bacteria in splay regions of the pattern exhibiting a transition from a rectilinear jet to an undulating jet as the local concentration of bacteria increases thanks to the swimmers joining the jet from the neighboring bend regions. Local concentration is calculated for each region of the jet, using its local width and contour length. **b,** Local director (red arrows) and bacteria body axes (blue arrows) for the jet in part **a**. **c,** Map of the angle $\beta$ between the bacteria and the pre-imposed director. **d,** Optical retardation and the apparent director field for the undulating instability as visualized by PolScope. The scale shows pseudocolors that correspond to the optical retardance of the cell. Low-retardance regions correspond to the twist of the director caused by bacteria in the undulating jet; in these regions, the director field cannot be determined accurately by PolScope.

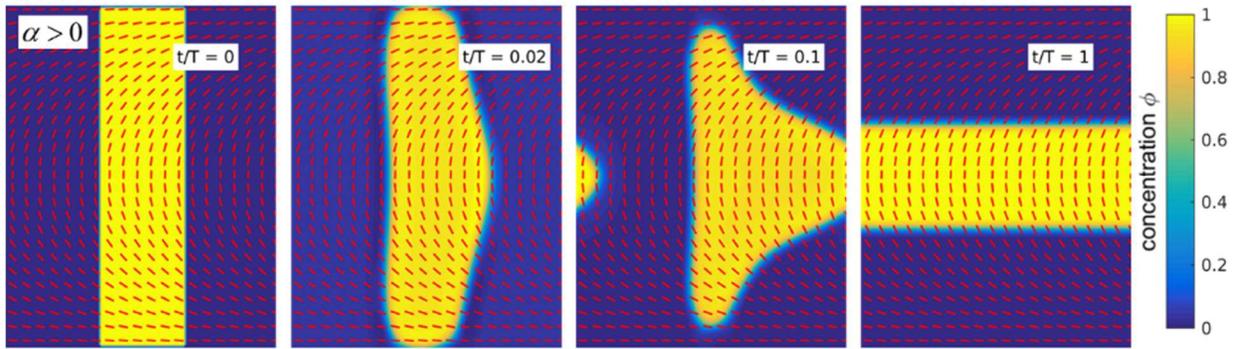

**Fig. S4.** Two-phase simulation of the contractile living nematic ( $\alpha > 0$ ). A blob of active nematic concentrates into bend regions for contractile activity, forming a jet along the positive $x$ direction. The high bacterial concentration is denoted by yellow; the depleted region is blue. Red solid lines mark the director field for the background liquid crystal orientation.

**Supplementary videos**

Video 1. Bacterial polar jets focused by the patterned "c" director field in the splay regions, moving from left to right. Bacteria moving in the opposite direction are realigned by experiencing a U-turn in bend regions. Contrast enhanced bright field microscopic images.

Video 2. Swimming bacteria in the uniformly aligned cell of 20 μm thickness with the gradient of bacterial concentration. The lower part of the sample with an elevated concentration of bacteria exhibits a bend instability; in the upper part, the swimmers follow the rectilinear trajectories imposed by the uniform director.

Video 3. Rectilinear jets experience undulations when the concentration of bacteria exceeds some threshold. The amplitude of undulations is stabilized by the underlying pre-imposed patterned director. Note that a small fraction of bacteria swim in the direction opposite to the jet. Contrast enhanced bright field microscopic images.

Video 4. Advection-diffusion simulation of bacterial jets and onset of their undulations. Left panel: variation of concentration of bacteria with time (shown by pseudocolors) and emergence of undulation as the concentration of bacteria increases; black ticks map the director of the passive nematic. Central and right panels: concentration (shown by pseudocolors) and total bacterial velocity fields in the lab frame (shown by black arrows) for $c^+$ and $c^-$ populations of bacteria.

Video 5. Two-phase simulation of jet formation by extensile swimmers in splay region. The high bacterial concentration is denoted by yellow; the depleted region is blue. The background liquid crystal orientation is shown by red solid lines.

Video 6. Two-phase simulation of jet formation by contractile swimmers in bend region. The high bacterial concentration is denoted by yellow; the depleted region is blue. The background liquid crystal orientation is shown by red solid lines.

Video 7. Two-phase simulation showing the undulating jet of extensile active fluid. The undulation is stabilized by the underlying passive liquid crystal, marked by red solid lines.

Video 8. Transport of single glass microspheres by the rectilinear bacterial jet moving from left to right in the splay region of the c-stripe patterned director.

Video 9. Transport of a chain of six glass microspheres by the rectilinear bacterial jet moving from left to right in the splay region of the c-stripe patterned director.

Video 10. Transport of micro-particle by an undulating bacterial jet.